\documentclass[aps,prl,showpacs,twocolumn,groupedaddress]{revtex4}
\usepackage{graphicx}
\newcommand{\vE} {{\bf E}}
\newcommand{\vB} {{\bf B}}
\newcommand{\vH} {{\bf H}}
\newcommand{\vD} {{\bf D}}

\newcommand{\ez} {{\bf e_z}}
\newcommand{\chir} {\chi_r}
\newcommand{\chii} {\chi_i}
\newcommand{\vk}   {{\bf k}}
\newcommand{\vecr} {{\bf r}}
\newcommand{\be} {\begin{equation}}
\newcommand{\ee} {\end{equation}}
\newcommand{\ba} {\begin{eqnarray}}
\newcommand{\ea} {\end{eqnarray}}

\begin{document}

\title{Universal transport properties of open microwave cavities
       with and without time-reversal symmetry}

\author{H. Schanze}
\affiliation{Fachbereich Physik der Philipps-Universit\"at Marburg,
D-35032 Marburg, Germany}

\author{H.-J. St\"ockmann}
\affiliation{Fachbereich Physik der Philipps-Universit\"at Marburg,
D-35032 Marburg, Germany}


\author{M. Mart\'\i nez-Mares}
\affiliation{Departamento de F\'{\i}sica, UAM-Iztapalapa,
Av. San Rafael Atlixco 186, Col. Vicentina,
09340 M\'exico D. F., M\'exico}
\affiliation{Instituto de F\'\i sica, Universidade do Estado do Rio de Janeiro,
R. S\~ao Francisco Xavier 524, 20550-900 Rio de Janeiro, Brazil}

\author{C. H. Lewenkopf}
\affiliation{Instituto de F\'\i sica, Universidade do Estado do Rio de Janeiro,
R. S\~ao Francisco Xavier 524, 20550-900 Rio de Janeiro, Brazil}


\date{\today}

\begin{abstract}
We measure the transmission through asymmetric and reflection-symmetric chaotic 
microwave cavities in dependence of the number of attached wave guides. Ferrite 
cylinders are placed inside the cavities to break time-reversal symmetry. The 
phase-breaking properties of the ferrite and its range of applicability are 
discussed in detail.  
Random matrix theory predictions for the distribution of transmission coefficients 
$T$ and their energy derivative $d T/d E$ are extended to account for absorption. 
Using the absorption strength as a fitting parameter, we find good agreement 
between universal transmission fluctuations predicted by theory and the experimental
data.
\end{abstract}

\pacs{05.45.Mt, 03.65.Nk, 73.23.-b}

\maketitle
\setcounter{secnumdepth}{1}

\section{Introduction}

There has been much theoretical interest in the universal 
transmission fluctuations through ballistic chaotic systems over 
the past years. This activity is partially driven by recent 
experiments on electronic conductance in open quantum dots. Random 
matrix theory was shown to be a valuable tool to obtain analytical 
results on the distribution of transmission and reflection 
coefficients, as well as on other related quantities \cite{Bee97}. 

Remarkably, there are very few ballistic experimental systems 
clearly showing universal transmission (or conductance) 
fluctuations as predicted by random-matrix theory. Conductance 
fluctuations in quantum dots \cite{Mar92} are already wanned by 
very small temperatures. Hence, theoretical clear-cut predictions 
of the transmission fluctuation dependence on the number of 
incoming and outgoing channels \cite{Bar94,Jal94} are hardly 
observed. Dephasing effects poses further difficulties 
\cite{Hui98b}, even considering that it can be incorporated into 
random matrix theory by introducing an additional 
phase-randomizing channel \cite{Bar95a}. Despite of this 
difficulties, quantum dots provided the first clear fingerprint of 
time-reversal symmetry breaking in the transmission distributions 
\cite{Hui98a}. Theory and experiment show an excellent agreement 
once the dephasing time is accounted for as a free parameter. 

An alternative to study universal transmission fluctuations is 
provided by microwave techniques. (There is a similarity to the 
conductante through quantum dots, that is proportional to the 
transmission - Landauer formula.) Transmission is directly 
measured in microwave experiments and cavities can be easily 
fabricated in any shape. Hence, this approach is ideally suited to 
verify theoretical predictions on transmission distributions. The 
first experiment of this type was performed by Doron {\it et al.} 
\cite{Dor90}. It may be considered as an experimental equivalent 
of the work by Jalabert {\sl et al.} \cite{Jal90} on conductance 
fluctuations in essentially the same system. The first, and up to 
now only study, aiming at the channel number dependence and the 
influence of time-reversal symmetry breaking is our own work 
\cite{Sch01d}. For the sake of completeness we would like to 
mention that there are two further microwave experiments on 
non-universal aspects of transmission \cite{Kim02}. 

Another quantity we shall examine in detail is the energy derivative of 
the transmission, $dT/dE$. The motivation stems from the study of the 
thermopower in electronic systems. There, one can show that the thermopower 
is proportional to the derivative of the conductance $G$ (or $T$) with 
respect to the Fermi energy (see, for instance, Ref. \cite{Lan98} for 
details and further references). 
Theory predicts a qualitative difference between diffusive and ballistic 
systems.
Whereas for a disordered wire the distribution of $dT/dE$ is expected to be 
a Lorentzian, for chaotic quantum dot systems one expects a distribution with 
a cusp at $E=0$.
This question has been addressed by a number of theoretical works 
\cite{Bro97c,Lan98,M-M03}.

The comparison between random-matrix-like fluctuations and microwave experiments 
has limitations.
It is not trivial to break time-reversal symmetry in
microwave systems. On the theoretical side, on the other hand,
analytical results are usually available for systems with broken
time-reversal symmetry only, whereas for systems with
time-reversal symmetry there are formidable technical problems.
One way to break time-reversal symmetry in microwave systems is to
introduce ferrites into the resonator \cite{So95,Sto95b}. In an
externally applied magnetic field the electrons in the material
perform a Larmor precession thus introducing a chirality into the
system, the precondition for breaking time-reversal symmetry.
It will become clear in what follows that this effect is unavoidably
accompanied by strong absorption. 

Thus, in microwave experiments there is either no time-reversal 
symmetry breaking, or strong absorption, or both.
Although meanwhile there is a number of works treating absorption
\cite{Kog00,Bee01b}, a better theoretical description of absorption
is still needed.

Last, but not least, the coupling between the cavity and the 
waveguides is usually not perfect (or ideal) in the experiments. 
Non ideal contacts mean that part of the incoming flux is promptly 
reflected at the entrance of the cavity and, hence, it is not 
resonant. (The same holds for quantum dots and leads.) While most 
theories assume ideal coupling, it is not difficult to account for 
non-ideal coupling \cite{Bro95b,Bee97}. The problem, however, is 
that to quantitatively determine the quality of the contacts, one 
needs to assess the phases of the $S$ matrix. This, in general, is 
not possible \cite{Men03}. We discuss this issue in our analysis. 

This paper is organized as follows. In Sec. \ref{sec2} we describe 
the experimental set-up and discuss how the addition of ferrite 
cylinders to the microwave cavities breaks time-reversal symmetry. 
The phase-breaking features of the ferrite and its absorption 
characteristics are discussed in detail in App. \ref{sec2.2}.
In Sec. \ref{sec3} we present the key elements of the statistical
theory for transmission fluctuations in ballistic systems.
Section \ref{sec:analysis} is devoted to the statistical analysis of 
our experimental data. 
We vastly expand an analysis of transmission fluctuations through 
asymmetric cavities previously presented \cite{Sch01d}.
Here we analyze new data on systems with reflection symmetry, where 
characteristic differences to systems without symmetry are expected 
\cite{Bar96a,Mar00}. We also discuss the distribution of the derivative
of the transmission with respect to energy, $dT/dE$. 
Our conclusions and an outlook of the open problems are presented in 
Sec.~\ref{sec:conclusions}.

\section{The experiment}
\label{sec2}

\begin{figure}
\includegraphics[width=\columnwidth]{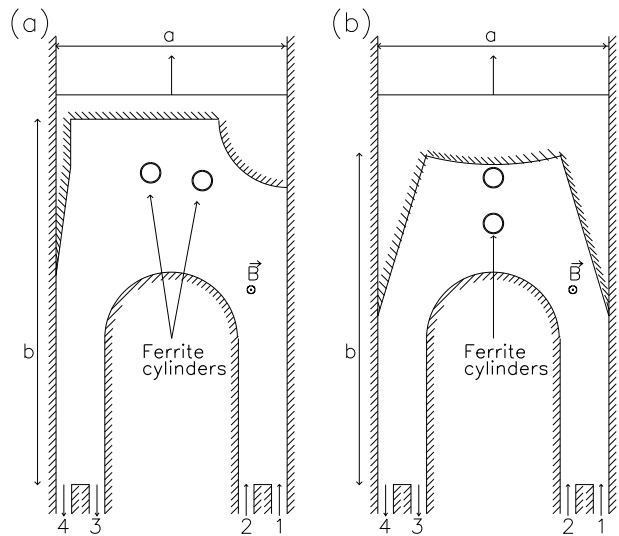}
\caption{Sketch of the microwave cavities used in the experiments.
(a) The asymmetric cavity has $a=237$\,mm and $b$ can vary from 375
to 425~mm. (b) The symmetric one has the same $a$, while $b$ ranges 
from 340 to 390~mm. The arrows indicate where the ferrite cylinders 
are placed. The entrance and exit waveguides are denoted by (1,2) and 
(3,4) respectively.}
\label{fig2.1}
\end{figure}

\begin{figure}
  \centering
  \includegraphics[width=\columnwidth]{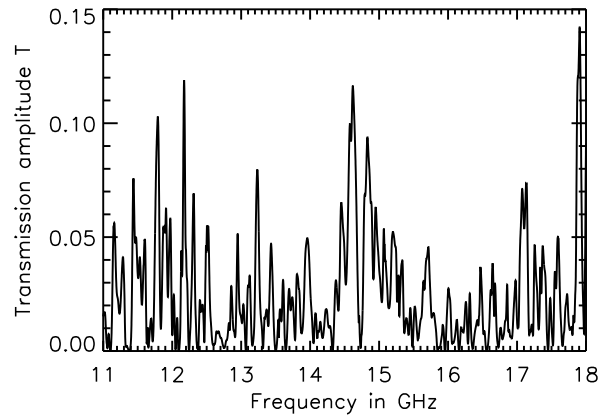}
  \caption{Typical transmission spectrum (asymmetric cavity).}
\label{fig2.2}
\end{figure}

Two different cavities were used in the experiment: an asymmetric 
and a symmetric one. Reflection symmetry is limited by the 
workshop precision. Figure \ref{fig2.1} displays their shapes. The 
height of cavities is $h=7.8$\,mm, i.\,e. both are 
quasi-two-dimensional for frequencies $\nu$ below $\nu_{\rm 
max}=c/2h=19.2$\,GHz. Two commercially available waveguides were 
attached both on the entrance and the exit side. The cut-off 
frequency for the first mode is at $\nu_1=c/2w=9.5$\,GHz where 
$w=15.8$\,mm is the width of the wave guides. Above 
$\nu_2=18.9$\,GHz a second mode becomes propagating. All 
measurements have been performed in the frequency regime where 
there is just a single propagating mode. The transmission 
coefficients were measured for all four possible combinations of 
entrance and exit waveguides. Figure \ref{fig2.2} shows a typical 
transmission spectrum. By varying the length $b$ of the resonator 
100 different spectra were taken, which were superimposed to 
improve statistics and to eliminate non-generic structures. A 
similar procedure has been already used in quantum dot experiments 
\cite{Hui98a,Hui98b}. 

We explore the ferrite reflection properties to break time-reversal
symmetry: We place two hollow ferrite cylinders, with radius $r=10$\,mm 
and thickness $d=1$\,mm inside the cavities.
The cylinders magnetization is varied by applying an external magnetic
field. At an induction of $B=0.475$\,T the ferromagnetic resonance is
centered at about $15.5$\,GHz.
The electrons in the ferrite perform a Larmor precession about the axis
of the magnetic field. At the Larmor frequency the ferromagnetic
resonance is excited giving rise to a strong microwave absorption.
This is, of coarse, unwanted.
Moving to frequencies located at the tails of the ferromagnetic resonance,
the microwaves are partially reflected and acquire a phase shift depending
on the sign of the propagation.
The ferrite cylinder has thus a similar effect on the photons as an
Aharonov-Bohm flux line in a corresponding electron system.
This correspondence has been already explored to study persistent currents
using a microwave-analog \cite{Vra02}.

This method to break time-reversal symmetry has an obvious and unavoidable
limitation: We have to move away from the ferromagnetic resonance frequency
to avoid strong absorption, but have to stay close enough to observe a
significant phase-breaking effect. In the present experiment, the optimal
frequencies occur on a quite narrow interval between 13.5 and 14.0\,GHz.

Appendix \ref{sec2.2} gives a quantitative description of the phase-breaking
mechanism due to the ferrite cylinders. Specific properties of the employed
ferrite, that are useful for the understanding of the experimental data, are
also discussed.


\section{Statistical theory}
\label{sec3}

There are two standard statistical theories that describe universal
transmission fluctuations of ballistic systems.
One is the $S$-matrix information-theoretical theory \cite{Mel99},
tailormade to calculate transmission distributions.
The other method, where the statistical $S$-matrix is obtained by
modeling the scattering region by a stochastic Hamiltonian \cite{Guh98},
is suited to the computation of energy and parametric transmission
correlation functions.
Both approaches were proven to be strictly equivalent in certain
limits \cite{Lew91}.
Complementing this result, there is numerical evidence supporting that
the equivalence is general \cite{Alves02}.
Here we use both methods: Our analytical results are obtained from the
information-theoretical approach, whereas the numerical simulations
are based on the stochastic Hamiltonian one.

We model the transmission flux deficit due to absorption by a
set of $N_\phi$ non transmitting channels coupled to the cavity.
We consider $N_1$ and $N_2$ propagating modes at the entrance and the
exit wave guides, respectively. The resulting scattering process is
described by the block structured $S$-matrix
\begin{eqnarray} \label{S-abs}
S = \left(\begin{array}{ccc}S_{1    1} & S_{1    2} & S_{1   \phi} \\
                            S_{2    1} & S_{2    2} & S_{2   \phi} \\
                            S_{\phi 1} & S_{\phi 2} & S_{\phi\phi} \end{array}\right)
  \equiv
    \left(\begin{array}{cc} \widetilde{S} &
          \begin{array}{r} S_{1   \phi} \\
                           S_{2   \phi} \end{array} \\
       \begin{array}{cc} S_{\phi 1} & S_{\phi 2} \end{array} & S_{\phi\phi} \end{array}
    \right)\;.
\end{eqnarray}
Here the set of indices $\{1\}$, $\{2\}$ label the $N_1$, $N_2$ propagating modes 
at the wave guides, while the set $\{\phi\}$ labels the $N_\phi$ absorption channels.
Transmission and reflection measurements, necessarily taken at the wave
guides, access directly only the $\widetilde{S}$ matrix elements.

Of particular experimental interest is the total transmission coefficient,
namely,
\be
T = \sum_{{a \in 2}\atop{b \in 1}} T_{ab}
\qquad \mbox{with} \qquad
T_{ab} \equiv |\widetilde{S}_{ab}|^2 \,.
\ee
The absorption at each $N_\phi$ channel can be quantified \cite{Lew92} by
$\Gamma_\phi = 1 - |\langle S_{\phi\phi} \rangle|^2$, where $\langle \cdots
\rangle$ indicates an ensemble average (described below).
We take the limits $N_\phi \rightarrow \infty$ and $\Gamma_\phi \rightarrow 0$,
while keeping $N_\phi \Gamma_\phi = \gamma$ constant.
In this way we mimic the absorption processes occurring over the entire
cavity surface, expressing their strength by a single parameter $\gamma$
\cite{Lew92}.
This modeling is equivalent to adding an imaginary part to the energy in the
$S$-matrix \cite{Bro97a}, a standard way to account for a finite $Q$-value
\cite{Dor90}.

We obtain the distributions $P_\beta(T)$ by numerical simulation.
To that end, we employ the Hamiltonian approach to the statistical $S$-matrix,
namely
\begin{equation}\label{SHapp}
S (E) = \mbox{$\openone$} - 2\pi i W^\dagger (E - H + i \pi W W^\dagger)^{-1} W \;,
\end{equation}
where $H$ is taken as a member of the Gaussian orthogonal 
(unitary) ensemble for the (broken) time-reversal symmetric case. 
This $S$ matrix parameterization is entirely equivalent to the 
$K$-matrix formulation recently used by Kogan and collaborators 
\cite{Kog00}. Since the $H$ matrix is statistically invariant 
under orthogonal ($\beta = 1$) or unitary ($\beta = 2$) 
transformations, the statistical properties of $S$ depend only on 
the mean resonance spacing $\Delta$, determined by $H$, and the 
traces of $W^\dagger W$. Maximizing the average transmission is 
equivalent to put tr$(W^\dagger W) = \Delta/\sqrt{\pi}$ 
\cite{Verbaarschot85}. This procedure can be used, in principle, 
to study any number $N$ of open channels. 

The simulations are straightforward: For every realization of $H$ 
we invert the propagator and compute $S(E)$ for energy values 
close to the center of the band, $E=0$, where the level density is 
approximately constant. The dimension of $H$ is fixed as $M=100 
\cdots 200$, depending on the number of channels $N$. The choice 
of $M$ represents the compromise between having a wide energy 
window for the statistics (large $M$) and fast computation (small 
$M$). For each value of $\gamma$ we obtain very good statistics 
with $10^4 \cdots 10^5$ realizations. 

We also analyze the fluctuations of the transmission coefficient 
energy derivative, $dT/dE$. We use the information-theoretical 
approach to analytically compute moments of $dT/dE$. For that 
purpose we express $dS/dE$ in terms of the $S$-matrix itself and a 
symmetrized form of the Wigner-Smith time-delay matrix $Q_E$ 
\cite{BrouwerPRL1997}, namely 
\begin{equation}
\label{eq:dSdE}
\frac{dS}{dE} = \frac{i}{\hbar} S^{1/2} Q_E S^{1/2} \, .
\end{equation}
Thanks to the well known statistical properties of $Q_E$ matrices, the computation
of $\langle (dT_{ab}/dE)^2 \rangle$ is possible \cite{M-M03}.
We note that Eq. (\ref{eq:dSdE}) is strictly valid only for $\Gamma_{\phi}=1$.
Hence, $\gamma$ is an integer number. Other values of $\gamma$ are obtained by 
extrapolation.

The full distribution of the transmission energy derivatives, $\widetilde{P}_{\beta}
(d T/d E)$, is obtained by numerical simulations. This is a simple extension of the 
numerical procedure described above. We compute $d S/d E$ directly from
\begin{equation} \label{deSHam}
\frac{d S}{d E} = 2\pi i W^\dagger (E - H + i \pi W W^\dagger)^{-2} W \;,
\end{equation}
at the same time as $S(E)$ is calculated.

Note that the only parameters of the theory are the mean resonance spacing $\Delta$, 
the number of channels $N$, and the absorption parameter $\gamma$. 
In what follows we analyze the cases of asymmetric and symmetric cavities.

\subsection{Asymmetric cavities}

To this point only stochasticity and orthogonal (time-reversal) or unitary 
(broken time-reversal) symmetry are assumed. Additional symmetries require 
special $S$-matrix parameterizations. Hence, the presented formalism is 
readily suited for asymmetric chaotic cavities.

\begin{figure}
  \centering
  \includegraphics[width=\columnwidth]{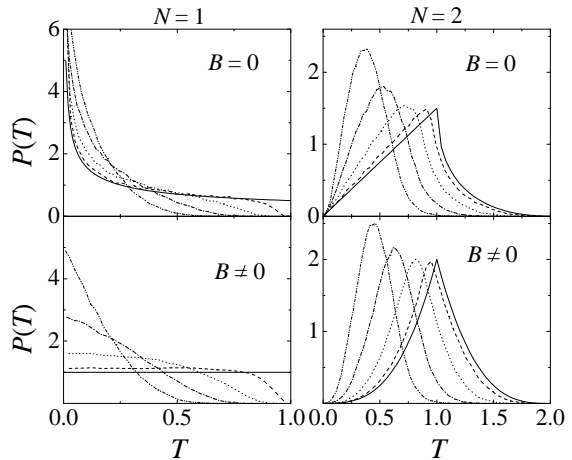}
  \caption{Transmission distribution $P(T)$ for asymmetric chaotic cavities with
           $N=1$ and $N=2$ open channels, both cases with ($B=0$) and without
           $(B\ne 0$) time-reversal symmetry. We consider different absorption
           parameters $\gamma$: 0 (solid), 0.25 (dash), 1 (dot), 2.5 (dash dot),
           and 5 (dash dot dot).
  }\label{fig3.1}
\end{figure}

Figure \ref{fig3.1} shows $P(T)$ for the $N=1$ and $N=2$ cases for various 
values of the absorption $\gamma$. One can nicely observe how the distributions 
for zero absorption \cite{Bar94} evolve to an exponential ($N=1$) or a convolution 
of exponentials ($N=2$) as the absorption strength $\gamma$ increases.
For $N=1$ our simulations are in excellent agreement with the analytical expression
obtained in Refs.~\cite{Sch01d,Bro97a}.
 
\begin{figure}
\centering
\includegraphics[width=\columnwidth]{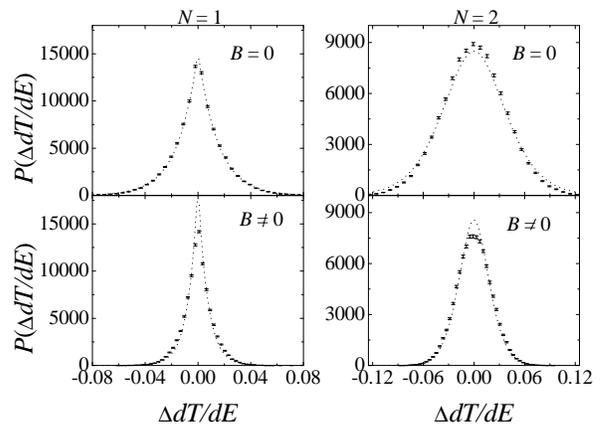}
\caption{Distributions of $d T/d E$ in units of inverse $\Delta$ for
asymmetric cavities. For $N=1$ the distributions agree with Eq. (\ref{PdeTab})
(dotted line), while for $N=2$ they follow Eq. (\ref{PdeT}) (dotted line).}
\label{wdetanum}
\end{figure}

For strong absorption, $\gamma \gg 1$, we find strong numerical evidence that
the distribution of individual channel-channel transmission energy derivatives,
$d T_{ab}/d E$, is exponential, namely
\begin{equation}\label{PdeTab}
\widetilde{P}_{\beta}(d T_{ab}/d E) = \frac{\lambda_{\beta}}{2}
\exp\left(-\lambda_{\beta} \left|\frac{d T_{ab}}{d E}\right|\right) \, ,
\end{equation}
where $\lambda_{\beta}$ depends on $\gamma$, but not the channel indices $a$ and
$b$. Furthermore, in this regime we find that the $dT_{ab}/dE$ for different pairs 
of channels are uncorrelated \cite{M-M03-2}.
We conclude that either this distribution is insensitive to dynamical channel-channel
correlations, or that such correlations are insignificant in our billiards.
Figure \ref{wdetanum} presents results for typical experimental values.
For independent $dT_{ab}/dE$, the distribution of $d T/d E$ for
$N=2$ is easily obtained by a convolution using Eq. (\ref{PdeTab}) and reads
\begin{eqnarray} \label{PdeT}
&& \widetilde{P}_{\beta}(d T/d E) = \frac{\lambda_{\beta}}{96}
\exp\left(-\lambda_{\beta} \left|\frac{d T}{d E}\right|\right)
\nonumber \\ && \times
\left( \lambda_{\beta}^3 \left|  \frac{d T}{d E} \right|^3 +
6 \lambda_{\beta}^2 \left|  \frac{d T}{d E} \right|^2 +
15 \lambda_{\beta} \left|  \frac{d T}{d E} \right| +
15 \right). \qquad
\end{eqnarray}

It remains to relate $\lambda_\beta$ to $\gamma$. This is done by computing 
$\langle (d T_{ab}/d E)^2\rangle$. The latter can be analytically
calculated using the energy derivative of the $S$-matrix, Eq. (\ref{eq:dSdE}), 
and reads \cite{M-M03-2}
\begin{eqnarray}
\left\langle \left( \frac{d T_{ab}}{d E} \right)^2 \right\rangle & = &
\frac{\pi^2}{\Delta^2} \frac{8}{\alpha^2(\alpha+1)^2} \nonumber \\
& \times & \frac{ \alpha^2 + \alpha - 2 + 4(2-\beta)}{ \alpha^2 +\alpha -2 
-4(2 -\beta)}\, ,
\end{eqnarray}
where $\alpha=N_1+N_2+\gamma$. Recalling that $\langle(d T_{ab}/d E)^2 
\rangle=2/\lambda_{\beta}^2$ we find $\lambda_\beta$ as a function of $\gamma$.

In Fig.~\ref{wdetanum} we compare the approximation $\widetilde{P}_\beta(d T/d E)$,
where $\lambda_\beta$ calculated as described above, with a direct numerical 
simulation. The agreement is rather good.

\subsection{Symmetric cavities}

The influence of absorption on the transmission fluctuations is even more 
pronounced in billiards with reflection symmetry. In the absence of absorption
the transmission distributions for reflection symmetric cavities was already
analytically computed.
The most salient features are the following:
When time-reversal symmetry is preserved, the theory predicts that the
transmission distribution $P(T)$ for reflection-symmetric cavities remains 
invariant when $T$ is substituted by $1-T$ \cite{GMM}.
On the other hand, for broken time-reversal symmetry, $P(T)$ coincides with 
the one for the {\it asymmetric} case, but with $T$ replaced by $1-T$ 
\cite{Bar96a}.

To fulfill the reflection symmetry, it is sufficient to consider the $S$-matrix with
the block structure \cite{GMM}
\begin{equation}
S = \left[ \begin{array}{cc}
\frac{1}{2}( S_1 + S_2 ) & \frac{1}{2}( S_1 - S_2 ) \\ & \\
\frac{1}{2}( S_1 - S_2 ) & \frac{1}{2}( S_1 + S_2 ) \end{array} \right] \, ,
\end{equation}
where $S_1$ and $S_2$ are unitary (and symmetric for $\beta=1$) $N_T/2 \times
N_T/2$ matrices with $N_T=2N+N_{\phi}$. Both $S_1$ and $S_2$ have the structure 
given by Eq.~(\ref{S-abs}).

The transmission coefficient now reads
\begin{equation}
T = \frac{1}{4} \sum_{a,b=1}^N
\big| [{S_1}]_{ab} - [{S_2}]_{ab} \big|^2 \equiv
\sum_{a,b=1}^N \sigma_{ab}\,.
\end{equation}

We numerically generate $P_\beta(T)$ and $\widetilde{P}_\beta(d T/d E)$ using the 
Hamiltonian approach to the $S$-matrix, Eqs. (\ref{SHapp}) and (\ref{eq:dSdE}). 
Now two statistically independent matrices, $S_1$ and $S_2$, are required. We 
chose the dimension of $H$ to be $M=50$. For each value of $\gamma$ we obtain 
very good statistics with $10^5$ realizations.

\begin{figure}
  \centering
  \includegraphics[width=\columnwidth]{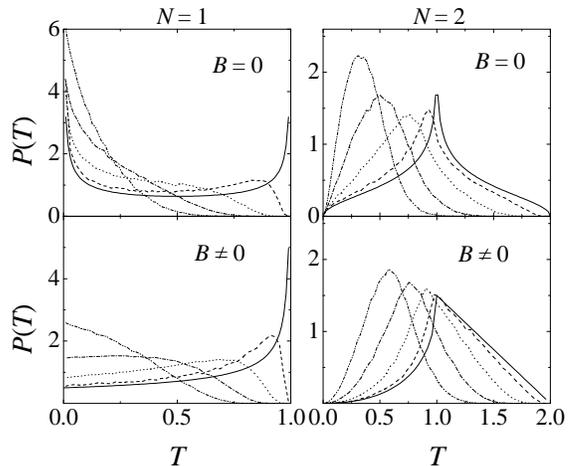}
  \caption{Symmetric cavity transmission distributions $P(T)$ for the one- and
           two-channel case. For $B=0$ we consider $\gamma = 0, 0.5, 2, 2.5,$ and
           10, corresponding to the solid, dashed, dotted, dashed-dotted,
           and dashed-dotted-dotted lines respectively. The same for the case of
           broken time-reversal symmetry, $B\neq 0$, but
           with $\gamma = 0, 0.25, 1, 2.5,$ and 5.}
\label{b1b2wtsnumeric}
\end{figure}

Figure \ref{b1b2wtsnumeric} contrasts $P_\beta(T)$ obtained analytically for
zero absorption \cite{Bar96a} with our numerical simulations for different
values of $\gamma$. Our analysis is restricted to the $N=1$ and $N=2$, as before.
We observe that with increasing $\gamma$ the fingerprints of the reflection symmetry
fade away, and the distributions become quite similar to those of asymmetric cavities.

As in the asymmetric case, for the strong absorption regime, $\gamma \gg 1$, our numerical
simulations strongly suggest that the distribution of the energy derivative of individual
channel-channel transmission coefficients $\widetilde{P}_\beta(d T_{ab}/d E)$ is exponential. 
However, in distinction to the asymmetric case, here the exponential law depends on the 
channels: The reflection symmetry (see Fig. \ref{fig2.1}) makes the channels (1,4) and (2,3) indistinguishable.
Accordingly, we find that the ``diagonal" coefficients $T_{14}$ and $T_{23}$, denoted
by $\sigma_{ab}^{\rm d}$, and the ``off-diagonal" ones $T_{24}$ and $T_{13}$, denoted
by $\sigma_{ab}^{\rm o}$ have different variance.
The second moment of the diagonal $d\sigma_{ab}^{\rm d}/d E$ is
\cite{M-M03-2}
\begin{eqnarray}
&& \left\langle \left( \frac{d\sigma^{\rm d}_{ab}}{d E} \right)^2
\right\rangle = \frac{\pi^2}{\Delta^2}
\frac{4}{(\alpha'-2){\alpha'}^2(\alpha'+1)^2} \nonumber  \\
&& \times \left[ \frac{ \alpha'(\alpha'-1)(7\beta-6)}{\alpha'+3} +
\frac{{(\alpha'}^2+\alpha'+2)(2-\beta)}{\alpha'+1}
\right] \qquad
\end{eqnarray}
whereas the off-diagonal is
\begin{eqnarray}
\left\langle \left( \frac{d\sigma_{ab}^{\rm o}}{d E} \right)^2
\right\rangle & = & \frac{\pi^2}{\Delta^2}
\frac{2({\alpha'}^2+\alpha'+2)}{(\alpha'-2){\alpha'}^2(\alpha'+1)^2(\alpha'+3)}
\nonumber \\
& \times & \left[ ( 2-\beta) \frac{\alpha'+2}{\alpha'+1} + 4(\beta-1) \right]. \qquad
\end{eqnarray}
Here, $\alpha'=\beta (N+\gamma/2)$.

\begin{figure}
\centering
\includegraphics[width=\columnwidth]{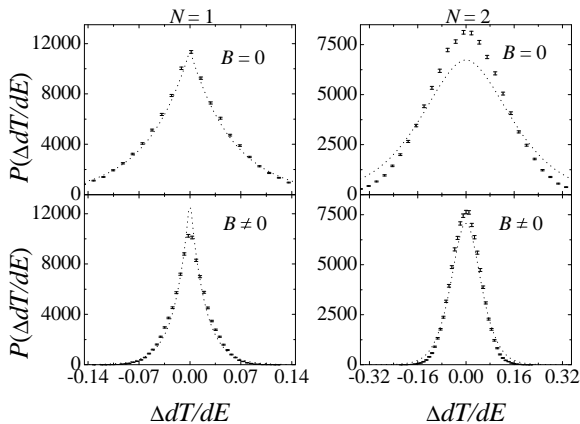}
\caption{Transmission energy derivative distributions for the symmetric chaotic
cavities. The points represent the results of the simulations for $\gamma=18$ (22) for
$B=0$ ($B\neq 0$) for $N=1$; $\gamma=14$ (18) for $B=0$ ($B\ne 0$) for
$N=2$. The dotted lines give the approximations (\ref{PdeTab}) and
(\ref{eq:PsymdTdEapprox}). For $N=1$ we present the diagonal case.}
\label{wdetsnum}
\end{figure}

For $\gamma \gg 1$, based on the numerical simulations,  we assume that 
$\widetilde{P}_\beta(d\sigma_{ab}/dE)$ is exponential and that for different
pair of channels $a$ and $b$ the $d\sigma_{ab}/d E$ are uncorrelated. 
We then equate $\mu_{\beta}^2=2/\langle (d\sigma_{ab}^{\rm d}/d E)^2\rangle$ 
and $\nu_{\beta}^2=2/\langle (d\sigma_{ab}^{\rm o}/d E)^2\rangle$ to write
\begin{eqnarray}
\label{eq:PsymdTdEapprox}
\widetilde{P}_\beta(d T/d E) & = & \frac{\mu_{\beta}^2\nu_{\beta}}{16}
\left[ \left( \frac{1}{\alpha_1} + \frac{1}{\alpha_2} \right)^2
\exp\left( -\frac{\nu_{\beta}}{2}
\left| \frac{d T}{d E}\right| \right)
\right. \nonumber \\ & + &
\left( \frac{1}{\alpha_1} - \frac{1}{\alpha_2} \right)
\left( \frac{1}{\alpha_1} + \frac{1}{\alpha_2} + \frac{1}{\mu_{\beta}} +
\left|\frac{d T}{d E}\right| \right)
\nonumber \\ & & \times \left.
\exp\left(-\mu_{\beta} \left|\frac{d T}{d E}\right| \right)
\right],
\end{eqnarray}
where $\alpha_1=\mu_{\beta}+\nu_{\beta}/2$, $\alpha_2=\mu_{\beta}-\nu_{\beta}/2$.
Figure \ref{wdetsnum} compares the approximation
$\widetilde{P}_\beta(d T/d E)$ with our numerical simulations. We chose
parameters realistic to out experiment. The agreement is quite good.
Deviations between the approximation (\ref{eq:PsymdTdEapprox}) and the numerical
simulations are of order $1/\gamma$.

\section{Statistical analysis of the experimental results}
\label{sec:analysis}

The statistical analysis of our experiment is based on two central hypothesis.
First, as standard, we assume that the transmission fluctuations of a chaotic
system are the same as those predicted by the random matrix theory \cite{Lew91}.
Second, we employ an ergodic hypothesis to justify that ensemble averages are 
equivalent to running averages, that is, averages over the energy (frequency) 
and/or shape parameters. This requires RMT to be ergodic  \cite{Pandey79}, 
which was recently shown \cite{Pluhar00}.
With few exceptions (see, for instance, Ref. \cite{Hemmady04}) this 
point is unnoticed. 

The experimental transmission coefficients were obtained by superimposing 100
different spectra measured for billiard lengths $b$ (see Fig.~\ref{fig2.1}).
In the studied frequency regime there is only a single propagating mode in 
each of the waveguides. Hence, to every waveguide we associate a single 
scattering channel.
For the $N=1$ case all measurements for the different combinations of entrance 
and exit waveguides were superimposed. The transmission for the $N=2$ case was 
obtained by combining the results from all $N=1$ measurements, namely,
$T=T_{13}+T_{14}+T_{23}+T_{24}$.

\begin{figure}
  \centering
  \includegraphics[width=\columnwidth]{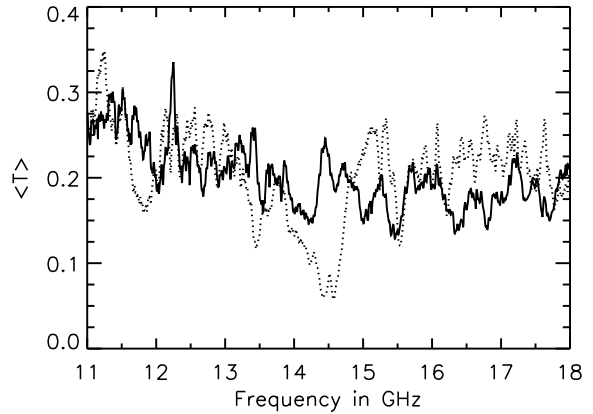}
  \caption{Mean Transmission $\langle T \rangle$ for the
           $N=2$ case for $B=0$ (solid line) and $B=0.470$\,T (dotted line).
          The Larmor resonance frequency is $\omega_R=2\pi \times 14.86\,$GHz) }
\label{fig4.0}
\end{figure}

\begin{figure}
  \centering
  \includegraphics[width=\columnwidth]{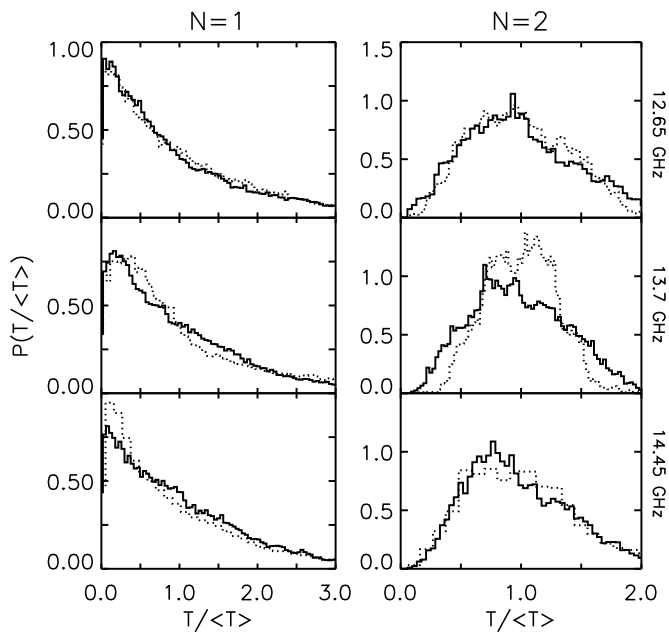}
  \caption{Transmission distribution for the $N=1$ (left) and $N=2$ (right)
           channel cases for three different frequency windows of width
           $\delta \nu = 0.3$\,GHz centered at $\nu_0$ (indicated in the figure).
          The histograms correspond to $B=0$ (solid line) and $B=0.470$\,T
         (dotted line). }
\label{fig4.1}
\end{figure}

Figure \ref{fig4.0} shows the mean transmission ($N=2$ case) with 
and without applied external magnetic field. When related to 
experimental quantities, $\langle \cdots \rangle$ indicate running 
averages. Using the Weyl formula, we associate the frequency $\nu$ 
(actually, $\nu/\Delta$) with the energy $E$ introduced in the 
preceding section. The strong absorption due to the Larmor 
resonance is clearly seen. In App. \ref{sec2.2} we discuss why is 
the phase-breaking effect expected to be best observed in the 
tails of the Larmor resonance. Figure \ref{fig4.1} illustrates 
this very nicely. It shows the scaled transmission distribution 
$P(T/\langle T \rangle)$ for the asymmetric billiard in three 
different frequency windows both with and without applied external 
magnetic field. It is only in the frequency interval form 13.55 to 
13.85\,GHz that $P(T/\langle T\rangle)$ changes with magnetic 
field. We stress that this is different from just an absorption 
effect: In the frequency window around $14.45$\,GHz, where the 
absorption is strongest, the normalized distributions with and 
without magnetic field are basically the same (the only difference 
is in the mean transmission). We identify the change in 
$P(T/\langle T\rangle)$ with the expected phase-breaking effect 
and assume that the applied magnetic field is sufficient for the 
ferrite cylinders to fully break time-reversal symmetry. Similar 
observations were made for the symmetric billiard. 

Before we present our statistical analysis, it remains to discuss how ideal
the cavity-waveguides coupling is.
To determine the antenna coupling we measured the transmission through 
two waveguides facing each other directly. In the whole applied frequency 
range the total transmission was unit, with an experimental uncertainty 
below $5\%$, showing that the antenna coupling is perfect. There are, 
however reflections of about $10\%$ in amplitude from the open ends of 
the waveguides, where they are attached to the billiard. 
Small deviations from ideal coupling are also consistent, for the frequencies 
we work, with Ref. \cite{Men03}. 
Since the absorption is strong in the present experiment, and an imperfect coupling 
can be compensated for to a large extent by a rescaled absorption constant, 
we decide not to explicitly account for coupling corrections.
In summary, throughout the forthcoming analysis we assume perfect coupling between 
the cavity and  the waveguides.

For the sake of clarity, we present the statistical analysis of the asymmetric
and the symmetric cavities separately.

\subsection{Asymmetric cavity distributions}

\begin{figure}
  \centering
  \includegraphics[width=\columnwidth]{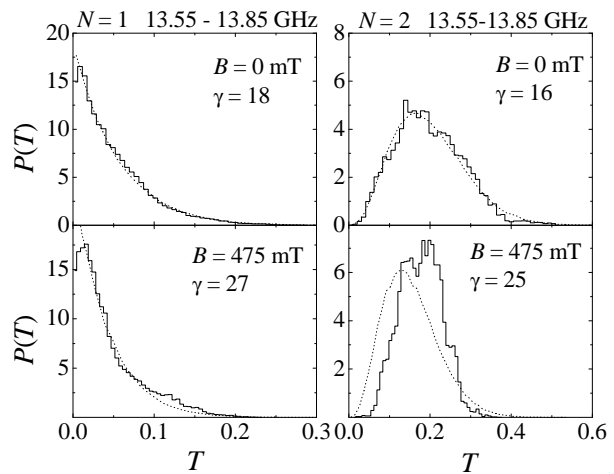}
  \caption{Transmission distributions for the asymmetric cavity. The histograms
           correspond to data taken within the indicated frequency window.
           The dotted lines stand for the random matrix simulations, with
           $\gamma$ as a fitting parameter.}\label{wta}
\end{figure}

Figure \ref{wta} compares the experimental transmission distributions in the
``phase-breaking" frequency window with the statistical theory. 
The absorption parameter $\gamma$, see Sec. \ref{sec3}, was adjusted to give the 
best fit of the theoretical $\langle T \rangle$ to the experiment. 
The agreement is excellent, except for $N=2$ with $B\neq 0$.

We work with a single asymmetric cavity, but use different $\gamma$ values
for $N=1$ and $N=2$.
The reason is simple: For $N=2$ we consider the contributions from all antennas
to the transmission, whereas for $N=1$ two antennas act as additional absorption
channels. This gives rise to a simple relation, namely, $\gamma^{(N=1)}=
\gamma^{(N=2)}+2$.

\begin{figure}
  \centering
  \includegraphics[width=\columnwidth]{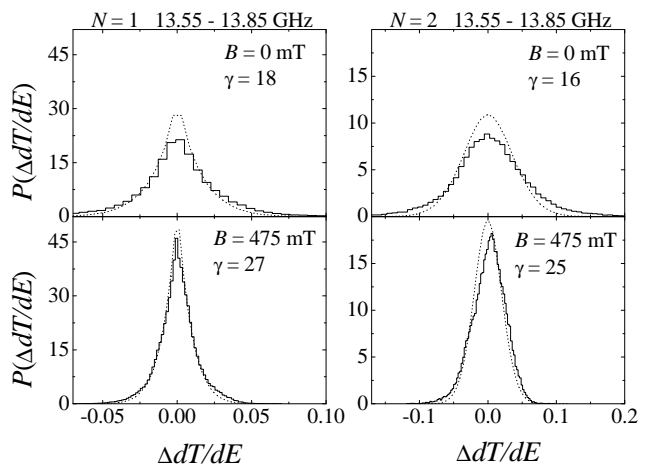}
  \caption{Distribution of the energy derivative of the transmission for the
           asymmetric cavity. The dotted lines correspond to the theoretical
           distributions.}
\label{wdeta}
\end{figure}

In order to compare the experimental transmission energy derivative distributions
with the universal random matrix results we have to rescale the experimental data
by the mean resonance spacing, namely,
$d T/d E \rightarrow \Delta d T/d E$. We use the Weyl
formula to estimate $\Delta$. Figure \ref{wdeta} shows a comparison between theoretical
and experimental results for $P(\Delta d T/d E)$. Note that we take the
same $\gamma$ as for $P(T)$. The signatures of the channel number, and the influence of
time-reversal symmetry breaking are clearly seen. We checked that the increase in
absorption when switching on the magnetic field, without switching to the unitary
ensemble as well, is not sufficient to reproduce the data. Inaccuracies in the
assessment of $\Delta$ provide a possible explanation for the slight disagreement
between theory and experiment. The Weyl formula does not account for the standing waves
in the ferrite cylinders and, thus, overestimates $\Delta$. This is consistent with
Fig. \ref{wdeta}.

\begin{figure}
  \centering
\includegraphics[width=\columnwidth]{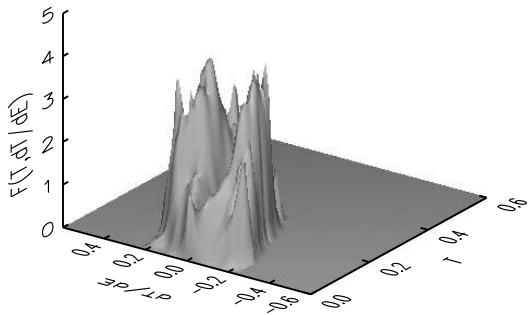}
  \caption{Normalized joint distribution $F(T,d T/d E)=P(T,d T/d E)/[P(T)P(d T/d E)]$
   for the asymmetric cavity for $N=1$, $B=0$. Similar result holds to $B\neq 0$.}
\label{b1n1fcor}
\end{figure}
\begin{figure}
  \centering
\includegraphics[width=\columnwidth]{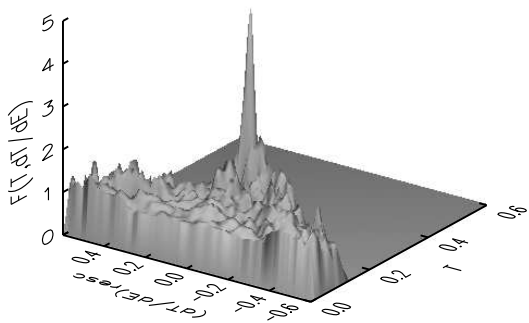}
  \caption{Same as in Fig. \ref{b1n1fcor}, but with $d E/d T$ replaced by
$(d T/d E)_{resc}=(d T/d E)/\sqrt{T(1-T)}$.}
\label{b1n1fcorresc}
\end{figure}

The joint distribution of $T$ and $d T/d E$ was studied in Ref.~\cite{Bro97c} for 
$N=1$ and $\gamma=0$. Remarkably, it was found that albeit $T$ and $d T/d E$ are 
correlated, the rescaled quantity $(d T/d E)_{\rm resc}=(d T/d E)/\sqrt{T(1-T)}$ 
and $T$ are not. We checked if this finding holds in our experiment, despite of
absorption.
Figure \ref{b1n1fcor} shows the ''normalized" joint probability $F(T,\Delta d T/d E)
\equiv P(T, \Delta dT/dE)/[P(T)P(\Delta dT/dE)]$ 
in a three dimensional representation for $N=1$ and $B=0$. 
A clear correlation is observed. To contrast, Fig. \ref{b1n1fcorresc} shows
$F[T,\Delta(d T/d E)_{\rm resc}]$. 
Here the distribution becomes flat.
Unfortunately we do not have enough statistics to make a reliable
determination of the distribution. A similar result, not shown here, 
holds for the $B\neq 0$ case.

\subsection{Symmetric cavity distributions}
\label{sec:symmetricdistributions}

We switch now to the statistical analysis of the symmetric cavity transmission
fluctuations.

\begin{figure}
  \centering
  \includegraphics[width=\columnwidth]{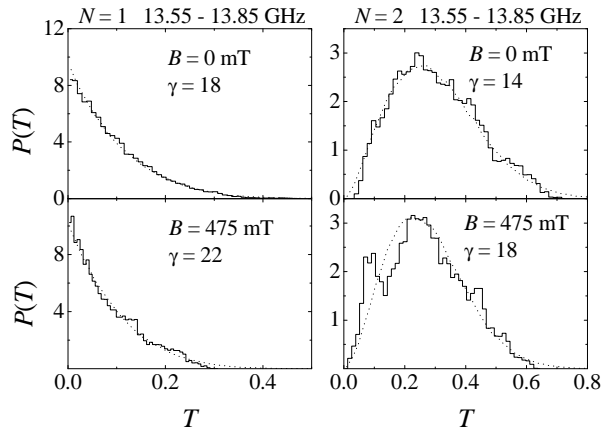}
  \caption{Transmission distributions $P(T)$ for the symmetric billiard. Histograms stand
           for the data taken at the indicated frequency interval, whereas the 
           dotted lines correspond to the simulations. The absorption 
           $\gamma$ is a fitting parameter.}
\label{b1b2wtsexp}
\end{figure}

Figure \ref{b1b2wtsexp} shows the experimental shows $P(T)$ for transmissions
within $13.55 \le \nu \le 13.85$ GHz, where the phase-breaking effect is expected
to be strongest. 
As before, the absorption parameter $\gamma$ is the best fit of the theory to 
the experiment. Here, for all studied cases a nearly perfect agreement is found.
Now $\gamma^{(N=1)}=\gamma^{(N=2)}+4$. This is due the reflection symmetry.


\begin{figure}
  \centering
\includegraphics[width=\columnwidth]{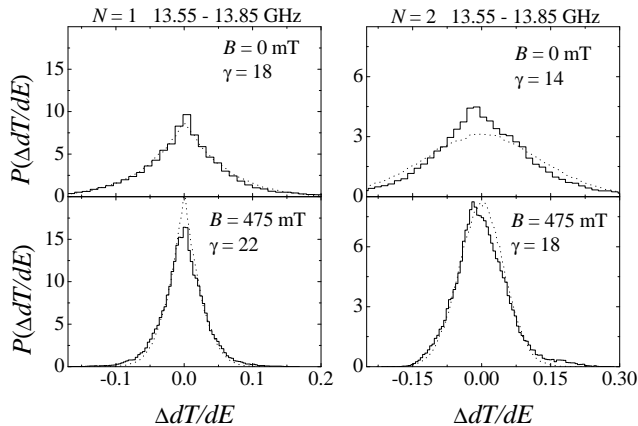}
  \caption{Distribution of the energy derivative of the transmission for the symmetric
cavity. The dotted lines corresponds to the theoretical distributions obtained from
random matrix theory.}
\label{wdets}
\end{figure}

Figure \ref{wdets} shows the experimental distributions $P(\Delta d T/d E)$ for 
the symmetric case. The signatures of the channel number and the influence of 
breaking time-reversal symmetry, are clearly seen. For all cases of the symmetric 
billiard the theoretical curves are plotted as well. We observe that the experimental
distributions verify the overall trends of the theoretical predictions.
In particular, the characteristic cusp at $E=0$ is nicely reproduced for $N=1$. 
Similar to the asymmetric case, the agreement between experiment and theory is not 
as good as for the transmission distribution.

\begin{figure}
  \centering
\includegraphics[width=\columnwidth]{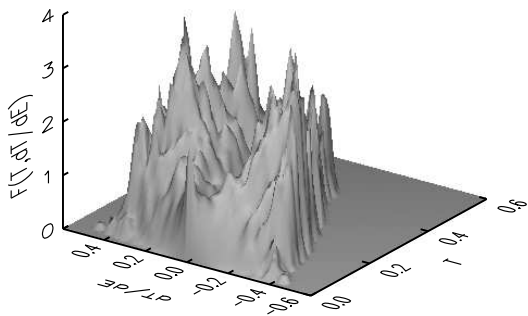}
  \caption{Normalized joint distribution
  $F(T,d T/d E)=P(T,d T/d E)/[P(T)P(d T/d E)]$
   for the symmetric cavity for $N=1$, $B=0$. Similar result holds to $B\neq 0$.}
\label{b1n1fcorsym}
\end{figure}
\begin{figure}
  \centering
\includegraphics[width=\columnwidth]{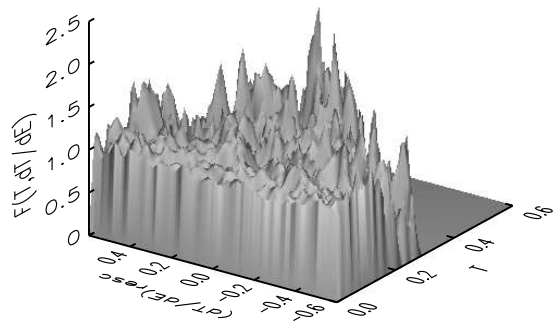}
  \caption{Same as in Fig. \ref{b1n1fcorsym}, but with $d E/d T$ replaced by
$(d T/d E)_{resc}=(d T/d E)/\sqrt{T(1-T)}$.}
\label{b1n1fcorrescsym}
\end{figure}

As in the case of asymmetric cavities, theoretical calculations 
\cite{M-M03} show that although $T$ and $d T/d E$ are correlated, 
the rescaled quantity $(d T/d E)_{\rm resc}=(d T/d 
E)/\sqrt{T(1-T)}$ is independent of $T$. Here also the analytical 
results were obtained for the $N=1$ case. Fig. \ref{b1n1fcorsym} 
shows the normalized joint probability $F(T,d T/d E)=P(T,d T/d 
E)/[P(T)P(d T/d E)]$ in a three dimensional representation for 
$N=1$, $B=0$ case. A clear correlation is manifest. For 
comparison, Fig. \ref{b1n1fcorrescsym} shows the corresponding 
quantity for $f[T,(d T/d E)_{\rm resc}]$. Now the correlation has 
vanished, in accordance with theory. Similar result, not shown 
here, holds for the $B\neq 0$ case. 

\section{Conclusions}
\label{sec:conclusions}

This work shows that microwaves are ideally suited to experimentally verify 
the theory of universal transmission fluctuations through chaotic cavities. 
The results presented in the present paper would have been hardly accessible 
by any other method.

We observe a nice overall agreement between our experimental data and the random
matrix results. However, the comparison between theory and microwave experiment is limited 
by the following issues.

In experiments, the coupling between waveguides and the cavity is 
usually not ideal, whereas in most theoretical works ideal 
coupling is assumed. In the frequency range we work \cite{Men03} 
supports our working hypothesis of nearly perfect coupling. In 
general, however, it turns out that without measuring the S matrix 
(with phases) it is hard to disentangle direct reflection at the 
cavity entrance (imperfect coupling) from absorption. From the 
experimental side, it would be desirable to have a better handle 
on absorption.

Microwave systems are usually time-reversal invariant, and as we have seen it is not 
trivial to break this symmetry. At the same time we increase the magnetic field, turning 
on the phase-breaking mechanism, absorption also increases. Unfortunately, both effects
are inextricable. This is why it is beyond our present experimental capability to 
quantitatively investigate the transmission fluctuations along the crossover regime between
preserved and broken time-reversal invariance.
Actually, to compare theory  with experimental results we assume that the transmission 
data at $B =0.470$ mT and $13.55 < \nu < 13.85$ GHz are far beyond the crossover regime.

We hope that the present work will trigger additional theoretical effort in the
mentioned directions.

\acknowledgments

C. W. J. Beenakker is thanked for numerous discussions at different stages of this
work. We also thank P. A. Mello for suggesting the symmetric cavities measurements.
The experiments were supported by the Deutsche Forschungsgemeinschaft. MMM was
supported by CLAF-CNPq (Brazil) and CHL by CNPq (Brazil).

\appendix
\section{Phase-breaking properties of the ferrite}
\label{sec2.2}

This appendix is devoted to the discussion of the ferromagnetic
resonance and the phase-breaking mechanism.
For that purpose we first quickly present some elements of the
well-established theory of microwave ferrites, see for instance,
Ref. \onlinecite{lax62}.

\begin{figure}
\includegraphics[width=\columnwidth]{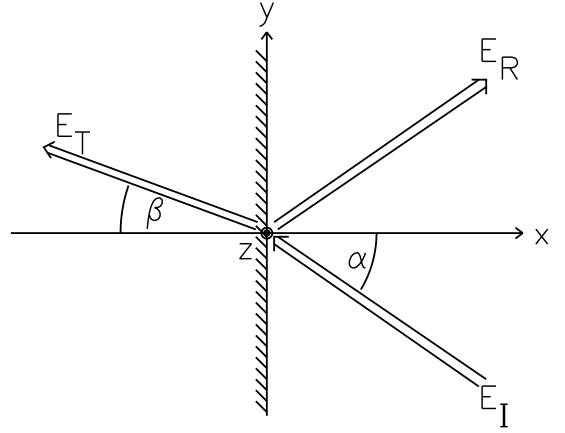}\\
\caption{\label{sketchref} Plane wave reflected by the surface of a ferrite slab.}
\end{figure}

For the sake of simplicity, we first restrict ourselves to the
situation of an incoming plane wave reflected by the surface of
an semi-infinite ferrite medium. We assume that incoming, reflected,
and refracted waves propagate in the $xy$ plane and are polarized
along the $z$ direction, and that there is an externally applied
static magnetic field in the $z$ direction, as shown in
Fig.~\ref{sketchref}.
We ask what is the phase acquired due to the reflection on the
ferrite.

To answer this question we need to solve Maxwell's equations.
For this geometry and single-frequency electromagnetic fields,
like our microwaves, this is a simple task.
The ferrite properties come into play by the constitutive
relations $\vD=\epsilon_0\epsilon\vE$ and $\vB=\mu_0\mu\vH$,
more specifically through the permeability $\mu$, that is a tensor
with the form
\begin{equation}\label{f10}
  \mu=1+\chi=\left(\begin{array}{ccc}
  1+\chir & -\imath\chii & \cdot \\
  \imath\chii & 1+\chir & \cdot \\
  \cdot & \cdot & 1+\chi_0
  \end{array}
\right)\,.
\end{equation}
with
\begin{equation}\label{f9}
    \chir=\frac{\omega_L\omega_M}
    {\omega_L^2-\hat{\omega}^2}\,,
    \quad \chii=-\frac{\hat{\omega}\omega_M}
    {\omega_L^2-\hat{\omega}^2}\,,
    \quad \hat{\omega}=\omega+\imath\lambda\,.
 \end{equation}
Here $\omega_L=-\gamma H_0$ and $\omega_M=\gamma M_0$ are the
precession angular frequencies about the external field $H_0$ and
the equilibrium magnetization $M_0$, respectively. $\mu_0$ is the
static susceptibility. More details can be found, for instance,
in Chapter 2.2.3 of Ref. \cite{Stoe99}.

We solve the proposed problem using for the electric field the ansatz
$\vE (r)=E(r)\ez$, where
\begin{equation}\label{f12}
  E(r)=\left\{\begin{array}{lc}
    E_Te^{\imath\vk_T \cdot \vecr}\,, & x<0\,, \\
    E_Ie^{\imath\vk_I \cdot \vecr}+E_Re^{\imath\vk_R\cdot\vecr}\,, & x>0\,, \
  \end{array}\right.
\end{equation}
with $\vk_I=k_0(-\cos\alpha,\sin\alpha\,,0)\,$,
$\vk_R=k_0(\cos\alpha,\sin\alpha\,,0)\,$, $\vk_T = k(-\cos\beta,\sin\beta\,,0)$,
see Fig.~\ref{sketchref}.

The derivation of the amplitudes $E_I$,$E_R$ and $E_T$ is similar
to that of Fresnel's formula (see, for instance, \cite{Jac62}).
Since an explicit calculation for ferrites is given in \cite{Vra02},
only the results shall be given. Using the continuity of
$\vE_\parallel$, $\vD_\perp$, $\vB_\perp$ and $\vH_\parallel$ on
the boundary, one writes
\begin{equation}
E_T=E_I+E_R \qquad \mbox{and} \qquad
k \sin \beta = k_0 \sin \alpha
\end{equation}
which is just Snell's law. For the relative amplitude of the
reflected part we obtain
\begin{equation}\label{f25}
\frac{E_R}{E_I}=\frac
  {(n^2/\epsilon)\cos\alpha+\imath\delta\sin\alpha-\sqrt{n^2-\sin^2\alpha}}
  {(n^2/\epsilon)\cos\alpha-\imath\delta\sin\alpha+\sqrt{n^2-\sin^2\alpha}}
\end{equation}
where
\begin{equation}
n^2=\frac{(\omega_L+\omega_M)^2-\hat{\omega}^2}{\omega_L(\omega_L+\omega_M)-\hat{\omega}^2}
\end{equation}
and
\begin{equation}\label{f26}
 \delta=\frac{\chii}{1+\chir}=-\frac{\hat{\omega}\omega_M}
  {\displaystyle\omega_L(\omega_L+\omega_M)-\hat{\omega}^2}\,.
\end{equation}
Note that there is a term depending on the sign of $\alpha$, i.e.
on the direction of the incident wave. This term is responsible
for the phase-breaking effect.

The above formulas have to be modified when dealing with a ferrite of 
finite width. For a slab of thickness $l$ and $\alpha=0$ we have
\begin{equation}\label{f28}
\frac{E_T}{E_I}=\frac{4\frac{\epsilon}{n}}{(1+\frac{\epsilon}{n})^2
e^{\imath k(1-n)l}-(1-\frac{\epsilon}{n})^2 e^{\imath k(1+n)l}}
\end{equation}
and
\begin{equation}\label{f29}
\frac{E_R}{E_I}=-2\imath \sin{knl}\frac{
1-\frac{\epsilon^2}{n^2}}{(1+\frac{\epsilon}{n})^2 e^{\imath
k(1-n)l}-(1-\frac{\epsilon}{n})^2 e^{\imath k(1+n)l}}\,.
\end{equation}
In contrast to Eq.\ (\ref{f25}), $E_T$ is no longer the amplitude of the
transmitted wave propagating inside the ferrite. Here $E_T$ is the amplitude
of the wave that crossed the ferrite slab an emerged at the other side.
The explicit formula for $\alpha \ne 0$ is lengthy and is not be presented
here.

\begin{figure}
\includegraphics[width=8cm]{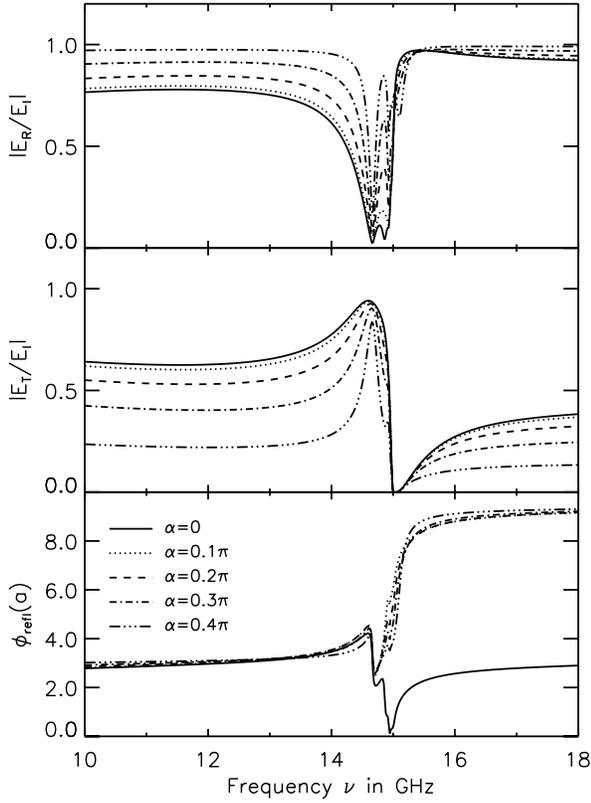}\\
\caption{\label{reflection} Reflection, transmission, and phase
shift for a ferrite slab ($M_0=$130 mT, $\epsilon=$15,
$\lambda=0.1\,$GHz) of thickness $l=$1 mm at $B_0=\mu_0H_0=$470
mT for different incidence angles $\alpha$.}
\end{figure}

The phase-breaking becomes clearly manifest by writing Eq.\ (\ref{f29})
as
\begin{equation}\label{f27}
\frac{E_R}{E_I}=\left|\frac{E_R}{E_I}\right|
 e^{i\phi_{\rm refl}(\alpha).}
\end{equation}
where $\phi_{\rm refl}(\alpha)$ is the phase acquired due to
reflection.
Figure \ref{reflection} shows modulus of transmission $|E_T/E_I|$
and reflection $|E_R/E_I|$ as well as the phase shift for
different incidence angles and $l=1\,$mm, the thickness of our
ferrite cylinders. The curves are calculated using the ferrite
parameters (see caption of Fig.~\ref{reflection}) given by the
supplier.
We find a resonance angular frequency of $\omega_R=\sqrt{\omega_L
(\omega_L+\omega_M)}=2\pi  \times 14.86\,$GHz.
This resonance corresponds to the dominant structure observed in
Fig.~\ref{reflection}. The additional substructures
are due to standing waves inside the ferrite.

\begin{figure}
\includegraphics[width=\columnwidth]{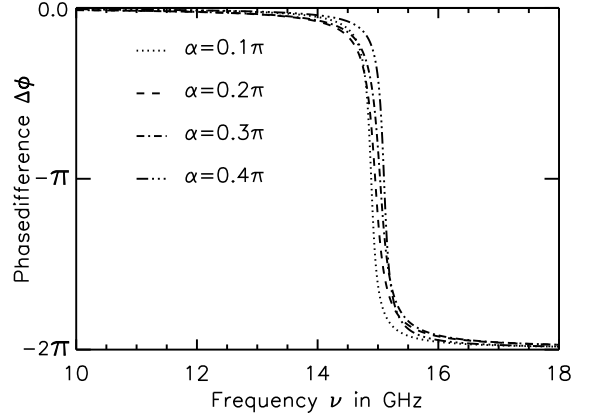}\\ 
\caption{\label{phase} 
Difference $\Delta\phi=\phi_{\rm refl}(\alpha)-\phi_{\rm refl}(-\alpha)$ of
phase shifts observed between an incoming wave and its time-reversed equivalent.}
\end{figure}

To illustrate the phase-breaking effect of the ferrite, in
Fig.~\ref{phase} we show the phase difference
$\Delta\phi=\phi_{\rm refl}(\alpha)-\phi_{\rm refl}(-\alpha)$
between the incoming and the time-reversed wave. We see that the
effect is maximal at the resonance frequency, and vanishes as one
moves off-resonance. Unfortunately, the absorption is maximal at 
the resonance too. This are the quantitative observations in
support of the discussion presented in Sec.~\ref{sec2}.

\begin{figure}
\includegraphics[width=8cm]{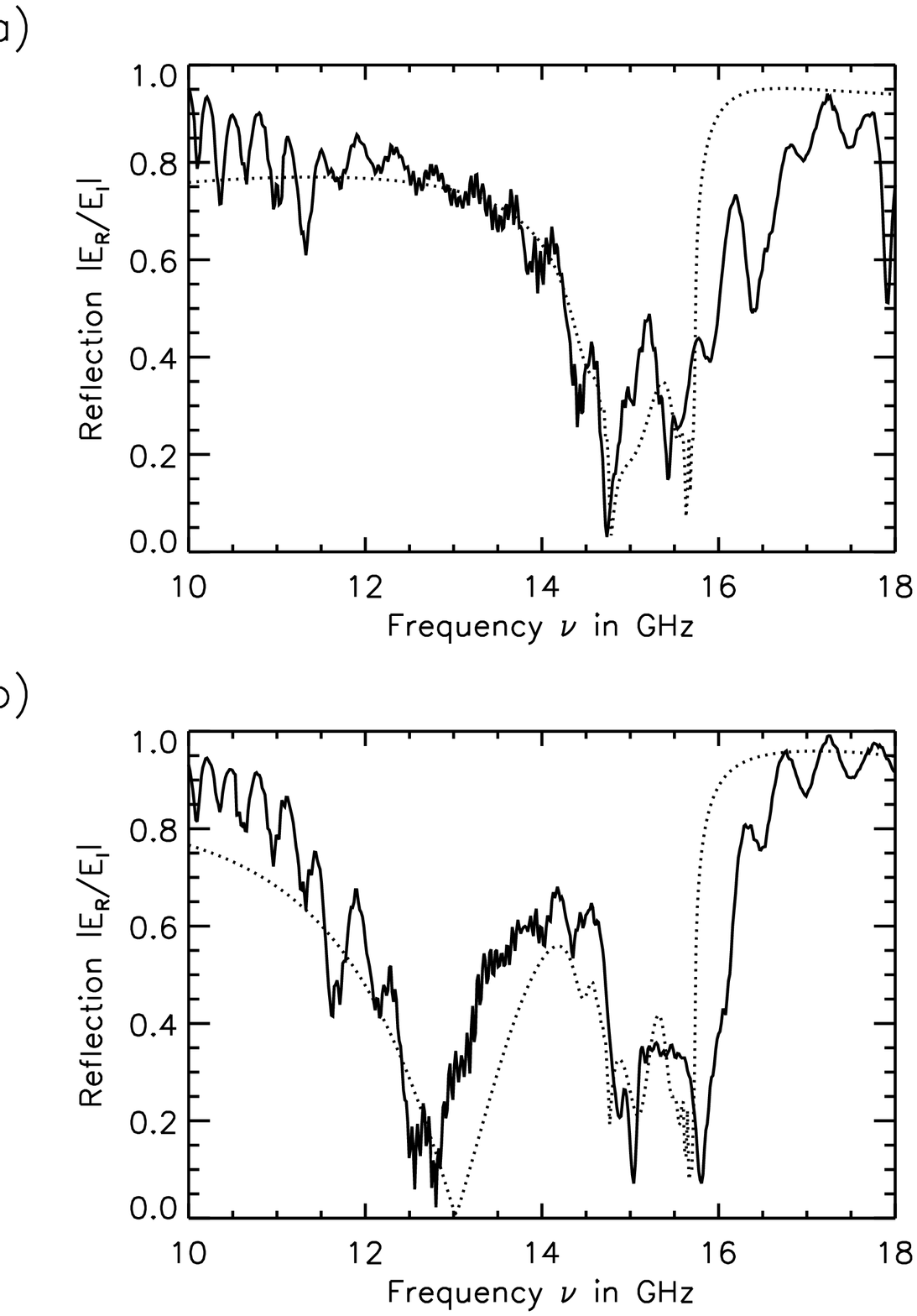}\\
\caption{\label{slabexperiment} Experimental reflection for a
ferrite slab of thickness $l=$1mm (a) and 2 mm (b). The dashed
lines have been calculated by superimposing the results for two
different internal magnetizations $M_0=$110 mT an 190 mT. The
broad minimum observed for $l=$2 mm close to 13 GHz is due to a
standing wave within the ferrite. For $l=$1 mm the corresponding
minimum is at 15 GHz and superimposes the ferromagnetic
resonance.}
\end{figure}

Finally, to experimentally check the properties of the ferrites, 
we place a small sheet of the material between two waveguide 
facing each other. Two different
thicknesses $l=1\,$mm and $2\,$mm were used. Figure \ref{slabexperiment} 
shows the measured reflection $|E_R/E_I|$ as a function of $\nu$. 
The small oscillations superimposing the dominant
resonance structures correspond to standing waves within the
waveguide and are an artifact of the experiment. Comparing the
experimental results with the calculation shown in Fig.~\ref{reflection}, 
we notice that the assumption of a single homogenous internal 
magnetization is not in accordance with the measurement.
The dashed line is obtained by superimposing the theoretical
results for two different values of the magnetization. The overall
behavior of the resonance structures becomes then
in qualitatively agreement with the data.


\end{document}